\documentstyle[12pt,aaspp]{article}    

\newcommand{\beq}{\begin{equation}}           
\newcommand{\eeq}{\end{equation}}             

\def\plotctyri#1#2#3#4{\centering \leavevmode
\epsfxsize=.45\columnwidth \epsfbox{#1}
\epsfxsize=.45\columnwidth \epsfbox{#2} \hfill
\epsfxsize=.45\columnwidth \epsfbox{#3}
\epsfxsize=.45\columnwidth \epsfbox{#4}}

\begin{document}

\title{Detecting Stellar Spots by Gravitational Microlensing}

\author{David Heyrovsk\'y and Dimitar Sasselov}

\affil{Department of Astronomy, Harvard University\\ 60 Garden St.,
Cambridge, MA 02138, USA}

\begin{abstract}

During microlensing events with a small impact parameter, the
amplification of the source flux is sensitive to the surface brightness
distribution of the source star. Such events provide a means for studying
the surface structure of target stars in the ongoing microlensing
surveys, most efficiently for giants in the Galactic bulge. In this work
we demonstrate the sensitivity of point-mass microlensing to small spots
with radii $r_s\lesssim0.2$ source radii. We compute the amplification
deviation from the light curve of a spotless source and explore its
dependence on lensing and spot parameters. During source-transit events
spots can cause deviations larger than 2\%, and thus be in principle
detectable. Maximum relative deviation usually occurs when the lens
directly crosses the spot. Its numerical value for a dark spot with
sufficient contrast is found to be roughly equal to the fractional radius
of the spot, i.e., up to 20\% in this study. Spots can also be
efficiently detected by the changes in sensitive spectral lines during
the event. Notably, the presence of a spot can mimic the effect of a
low-mass companion of the lens in some events.

\end{abstract}

\bigskip
\noindent
\bigskip
 
\centerline{To appear in {\it The Astrophysical Journal}, January 2000}
 
\clearpage 
\section{Introduction}
 
Apart from having a venerable history (Schwarzschild~1975), the question
of small-scale surface structure in normal stars is very important for
stellar modeling. Direct interferometric evidence is scarce and
inconclusive (Di Benedetto \& Bonneau~1990; Hummel et~al.~1994). Indirect
evidence, such as Doppler imaging and photometry, is limited to specific
types of stars (RS CVn, BY Dra, etc.). Photometric evidence for stellar
spots comes from the Optical Gravitational Lensing Experiment (OGLE)
survey of bulge giants (Udalski et~al.~1995). Modeling of the spots on
the stars selected from the OGLE database was reported by Guinan
et~al.~(1997). Direct evidence for a bright spot is available for one red
supergiant, $\alpha$ Ori (Uitenbroek, Dupree, \& Gilliland~1998),
although in this case the bright region may be associated with the
long-period pulsation of the star. The evidence for spots is particularly
scarce for normal red giants. Such information will be invaluable, given
the current difficulty in calculating detailed red giant atmosphere
models directly or extrapolating their physics from available models of
red dwarfs.

The presence of spots can also be revealed by observations of Galactic
microlensing events. During such an event the flux from a background star
is amplified by gravitational lensing due to a massive object, such as a
dim star, passing in the foreground. Over 350 such events have been
observed already, primarily toward source stars in the Galactic bulge and
the Magellanic Clouds. For a review of Galactic microlensing see
Paczy\'nski~(1996). In events with a small impact parameter, when the
lens passes within a few stellar radii of the line of sight toward the
source, the lens resolves the surface of the background star, as the
amplification depends on its surface brightness distribution. Various
aspects of this effect have been studied by Simmons, Newsam, \&
Willis~(1995), Gould \& Welch~(1996), Bogdanov \& Cherepashchuk~(1996),
Heyrovsk\'y \& Loeb~(1997)  and Valls-Gabaud~(1998), to name a few.
Spectral changes due to microlensing of a spotless source star were
described in detail recently by Heyrovsk\'y, Sasselov, \& Loeb~(1999;
hereafter HSL). In this complementary work we study the case when the
circular symmetry of the surface brightness distribution of the source is
perturbed by the presence of a spot. We illustrate the effect using a
model of a source similar to the one lensed in the MACHO Alert 95-30
event (hereafter M95-30; see Alcock et~al.~1997). During this
microlensing event the lens directly transited a red giant in the
Galactic bulge. As red giants in the bulge are generally the most likely
sources to be resolved, Galactic microlensing appears to be ideally
suited for filling a gap in our understanding of their atmospheres as
well as of spots in general.

In the following section we study the broadband photometric effect of a
spot on the light curve of a microlensing event. The spectroscopic
signature on temperature-sensitive lines is illustrated in \S 3. In \S 4
we discuss the limitations of the simple spot model used here, as well as
the possibility of confusing the effect of a spot with a planetary
microlensing event. Our main results are summarized in \S 5.

\section{Effect of a Spot on Microlensing Light Curves}

We describe the lensing geometry in terms of angular displacements in the
plane of the sky. The angular radius of the source star serves as a
distance unit in this two-dimensional description.

During a point-mass microlensing event the flux from a spotless star with
a limb-darkening profile $B(r)$ is amplified by a factor
\beq
A_{\ast}(\vec{r}_L)=\frac{\int B(r) A_0(|\vec{r}-\vec{r}_L|)\,d\Sigma}
{\int B(r)\,d\Sigma}\quad.
\label{eq:starampl}
\eeq
Here $\vec{r}_L$ is the displacement of the lens from the source center,
$\vec{r}$ is the position vector of a point on the projected surface of
the star $\Sigma$, and the point-source amplification is
\beq
A_0(\sigma)=\frac{\sigma^{2}+2\epsilon^{2}} 
{\sigma\sqrt{\sigma^{2}+4\epsilon^{2}}} \quad .
\label{eq:ptampl} 
\eeq
The Einstein radius of the lens\,\footnote{\, Also given in source radius
units.} is denoted by $\epsilon$; the separation between a point-source
and the lens is $\sigma$.

At large separations $\vec{r}_L$, formula~(\ref{eq:starampl}) is well
approximated by the point-source limit~(\ref{eq:ptampl}) with
$\sigma=r_L$. However, when the lens approaches the source closer than
three source radii, finite-source effects become observable (higher than
1\%--2\%), as shown in HSL. The light curve of such an event then
contains information on the surface structure of the source, introduced
through its dependence on the surface brightness $B(\vec{r}\,)$. HSL
dealt with spectral effects due to the wavelength dependence of $B(r)$ in
the case of a spotless source. Here we study the case of a source in
which the circular symmetry is perturbed by a spot.

In such a case it is useful to separate the surface brightness
distribution into the circularly symmetric component $B(r)$ (describing
the source in the absence of the spot) and a brightness decrement
$B_D(\vec{r}\,)$. This decrement is zero beyond the area of the spot
$\Sigma'$. The amplification of the spotted star can now be written 
\beq
A(\vec{r}_L)=\frac{\int B(r) A_0(|\vec{r}-\vec{r}_L|)\,d\Sigma\;- \int
B_D(\vec{r}\,') A_0(|\vec{r}\,'-\vec{r}_L|)\,d\Sigma'} {\int
(B-B_D)(\vec{r}\,)\,d\Sigma}\quad,
\label{eq:spotampl}
\eeq
where we made use of the linearity of the integral in the numerator in
$B$. Note that the second integral in the numerator is taken only over
the area of the spot. The position vector $\vec{r}\,'$ of a point within
the spot as defined here originates at the source center. The relative
deviation of the amplification from the amplification of a spotless 
source is
\beq
\delta (\vec{r}_L) = \frac{A-A_\ast}{A_\ast}(\vec{r}_L)=\frac{\int
B_D(\vec{r}\,')\,d\Sigma'} {\int(B-B_D)(\vec{r}\,)\,d\Sigma} \left[1-
\frac{\int B_D(\vec{r}\,') A_0(|\vec{r}\,'-\vec{r}_L|)\,d\Sigma'}
{A_\ast(\vec{r}_L)\,\int B_D(\vec{r}\,')\,d\Sigma'} \right] \quad .
\label{eq:delta} 
\eeq

We employ a simple model for a small spot by using a constant decrement
$B_D$ over a circular area with radius $r_s\ll 1$ on the {\em projected}
face of the source star, centered at a distance $s$ from its
center\,\footnote{\, Note that in this model the surface brightness within
the spot is not constant, it decreases toward the limb of the source
following the shape of the spotless profile $B(r)$.}. This way we can
compute both integrals in the numerator of equation~(\ref{eq:spotampl})
using the method for computing light curves of circularly symmetric
sources described in Heyrovsk\'y \& Loeb~(1997). The expression for the
amplification deviation in this model simplifies to
\beq
\delta(\vec{r}_L)=\frac{\pi r_s^2 B_D}{\int(B-B_D)(\vec{r}\,)\,d\Sigma}
\left[1-\frac{\int A_0(|\vec{r}\,'-\vec{r}_L|)\,d\Sigma'} {\pi r_s^2
A_\ast(\vec{r}_L)} \right]\quad .
\label{eq:dev}
\eeq

In the following we illustrate the effect of a test spot superimposed on a
spotless model of a red giant similar to the M95-30 source. The 3750 K
model atmosphere applied here is described in more detail in HSL. In most
of the applications we use its $V$-band limb-darkening profile for the
surface brightness distribution $B(r)$. The presented broadband results,
however, do not change {\em qualitatively} for different spectral bands
or source models in the small spot regime explored here.

The value of the brightness decrement $B_D$ depends on the physical
properties of the spot, as well as its apparent position on the source
disk. We can separate these two by expressing the decrement as $B_D =
(1-\mu^{-1})B(s)$. Here we introduced the contrast parameter $\mu$, which
is equal to the ratio of the spotless brightness at the position of the
center of the spot $B(s)$ to the actual brightness at the same position
$B(s)-B_D$. The dependence of $\mu$ on the spot position $s$ is weak
except when very close to the limb; we therefore neglect it here. The
parameter $\mu$ now depends purely on intrinsic physical properties -
namely, on the temperature contrast of the spot $\Delta$T, the effective
temperature of the star, and the spectral band. Its value can be obtained
numerically by comparing model atmospheres of different temperatures. For
example, $\mu\simeq 10$ for a $\Delta$T=1000 K spot on the model source
used. Values $\mu<1$ can be used to describe bright spots, while $\mu=1$
corresponds to zero contrast.

\begin{figure}[tb]
\plotone{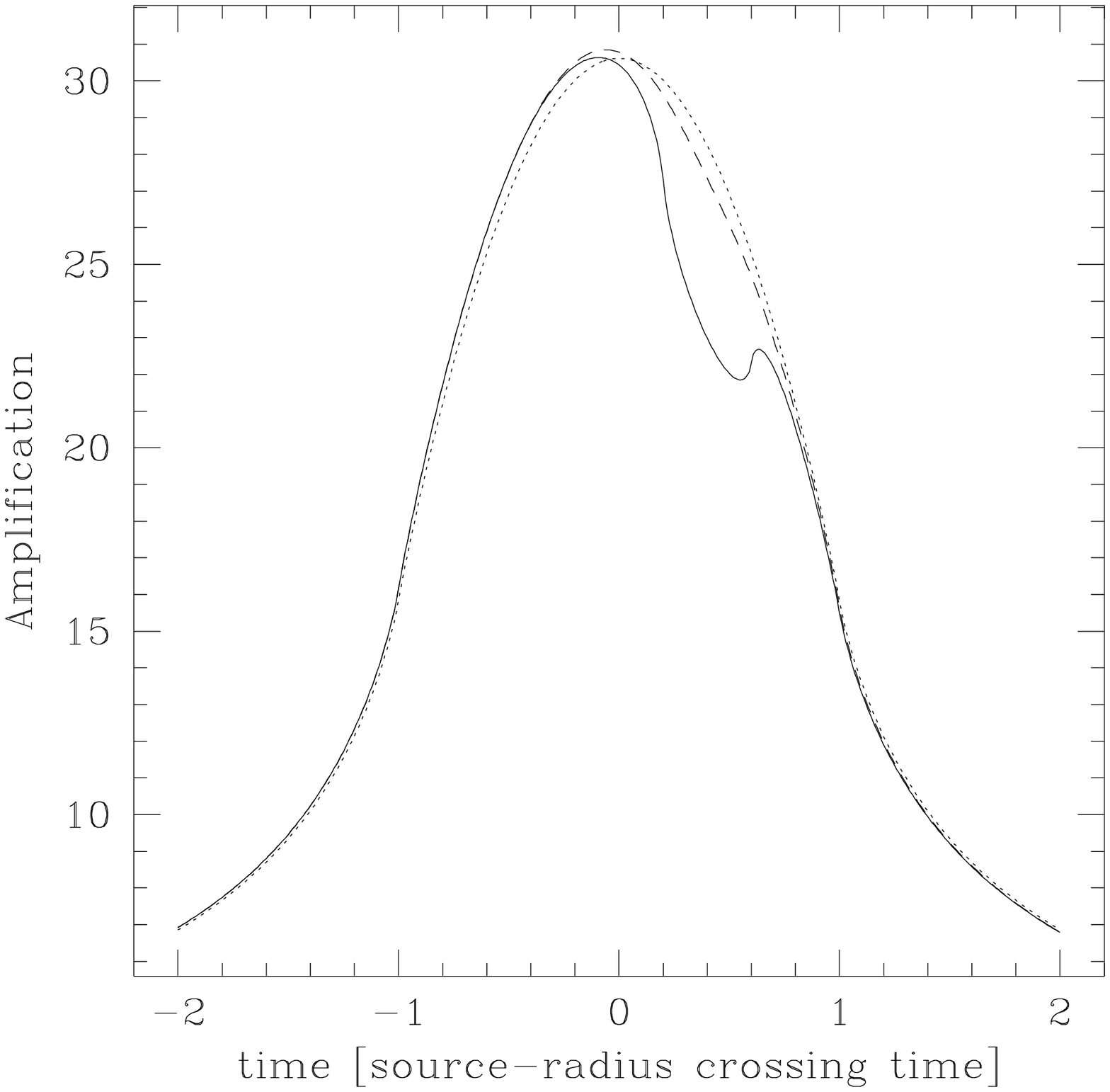}
\caption{Microlensing light curves of a star with a spot (lens with
$\epsilon=13.23$, zero impact parameter). {\em Solid line}: spot with
radius 0.2 centered at $s=0.4$ on the lens path; {\em dashed line}: same
spot offset by 0.3 perpendicular to lens path; {\em dotted line}: no spot
(for comparison). The underlying source is a 3750 K red giant observed in
the $V$-band, spot contrast $\mu=10$.}
\label{fig:spotlight}
\end{figure}

Sample light curves computed using equation~(\ref{eq:spotampl}) are
presented in Figure~\ref{fig:spotlight}. In all cases a lens with the
M95-30 Einstein radius of $\epsilon=13.23$ transits the source star with
zero impact parameter. The spot has a radius $r_s=0.2$ and contrast
$\mu=10$. When the lens directly crosses the spot ({\em solid line}; spot
centered at $s=0.4$), there is a significant dip in the light curve. On
the other hand, if the spot lies further from the lens path ({\em dashed
line}; closest approach to spot = 0.3, $s=0.5$), the effect is weak. It
consists primarily of a slight shift due to the offset center of
brightness, and a minor increase in peak amplification.

\begin{figure}[tb]
\plotone{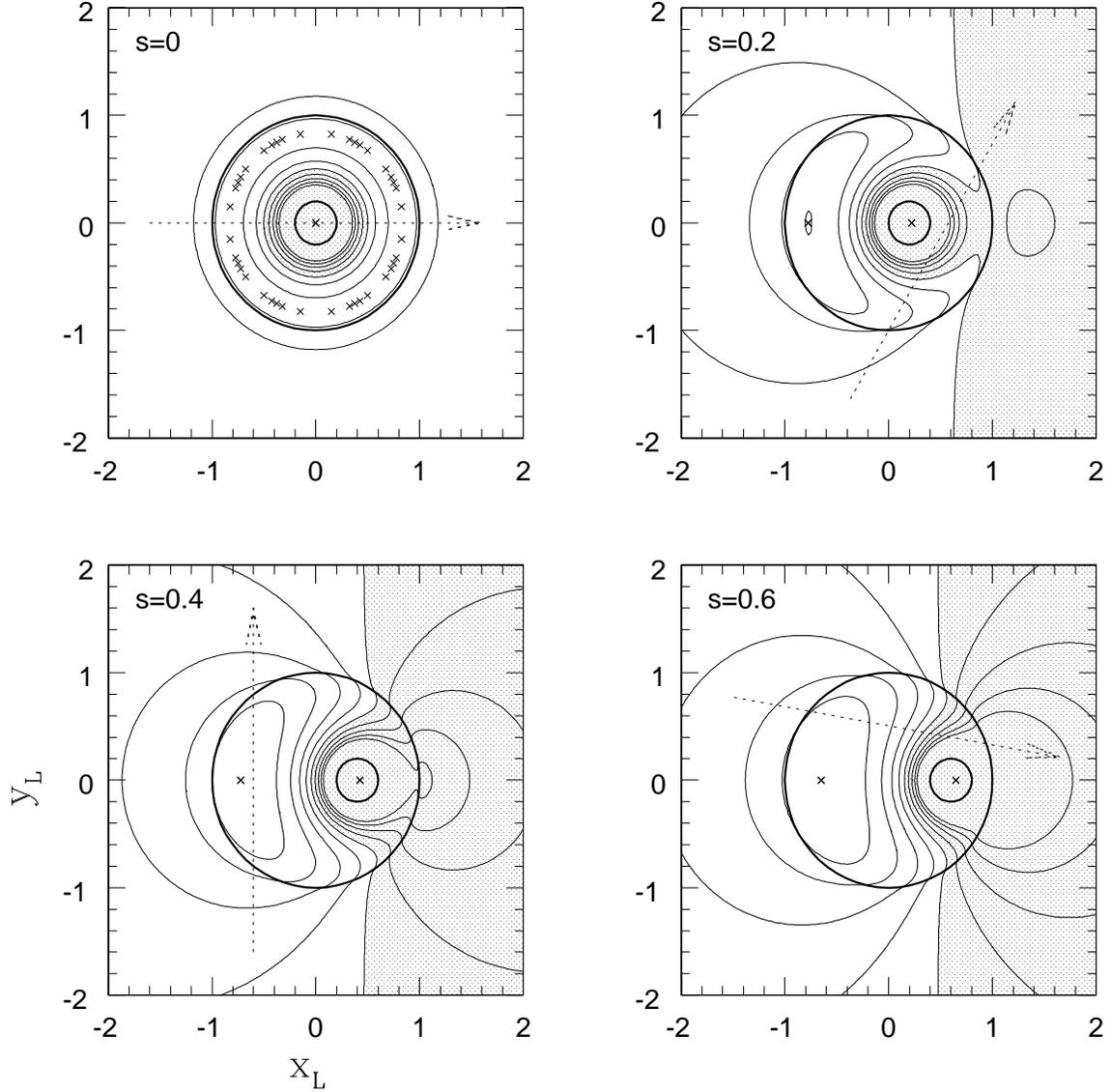}
\caption{Contour plots of relative amplification deviation $\delta$ as a
function of lens position (Einstein radius $\epsilon=13.23$). The two bold
circles in each plot represent the source with an $r_s=0.2$, $\mu=10$
spot. The four plots correspond to different spot positions, $s=0, 0.2,
0.4, 0.6$. The crosses identify the positions with maximum negative and
positive effects; the region with a negative effect is shaded. In the
$s=0$ case the positive maximum is extended along a circle. Contours are
spaced by 0.5\% decreasing toward the spot and increasing toward the
positive maximum. The minimum contour plotted here is -2\%; all positive
contours are plotted. Deviation curves for the lens paths marked by dotted
arrows are shown in Fig.~\ref{fig:devcurve}.} 
\label{fig:devcontour}
\end{figure}

To explore the range of possible spot signatures on light curves we study
the relative amplification deviation using equation~(\ref{eq:dev}). The
deviation, $\delta$, is primarily a function of parameters describing the
lensing and spot geometries ($\vec{r}_L, \epsilon, s, r_s$), and the spot
contrast $\mu$ - six parameters in all. The dependence of $\delta$ on the
lens position $\vec{r}_L$ is illustrated by the contour plots in
Figure~\ref{fig:devcontour}, for different spot positions $s$. The three
other parameters are kept fixed at values $\epsilon=13.23, r_s=0.2$ and
$\mu=10$. First we note that the deviation from the spotless light curve
of the same source can be positive as well as negative, for any spot
position\,\footnote{\,A dark spot can produce a positive effect, because
the amplification~(\ref{eq:spotampl}) is normalized by the lower
intrinsic flux from the spotted star.}. The negative effect peaks at
18\%--19\% at the spot position in all four cases. This region of the
source is relatively dimmer than in the spotless case. The weaker
positive effect (2\%--3\%, less when $s=0$) peaks on the opposite side of
the source close to the limb, a region relatively brighter than in the
spotless case. Geometrically the actual deviation depends on the
interplay of the distances of the lens from the spot, from the positive
peak and from the limb.

\begin{figure}[tb]
\plotone{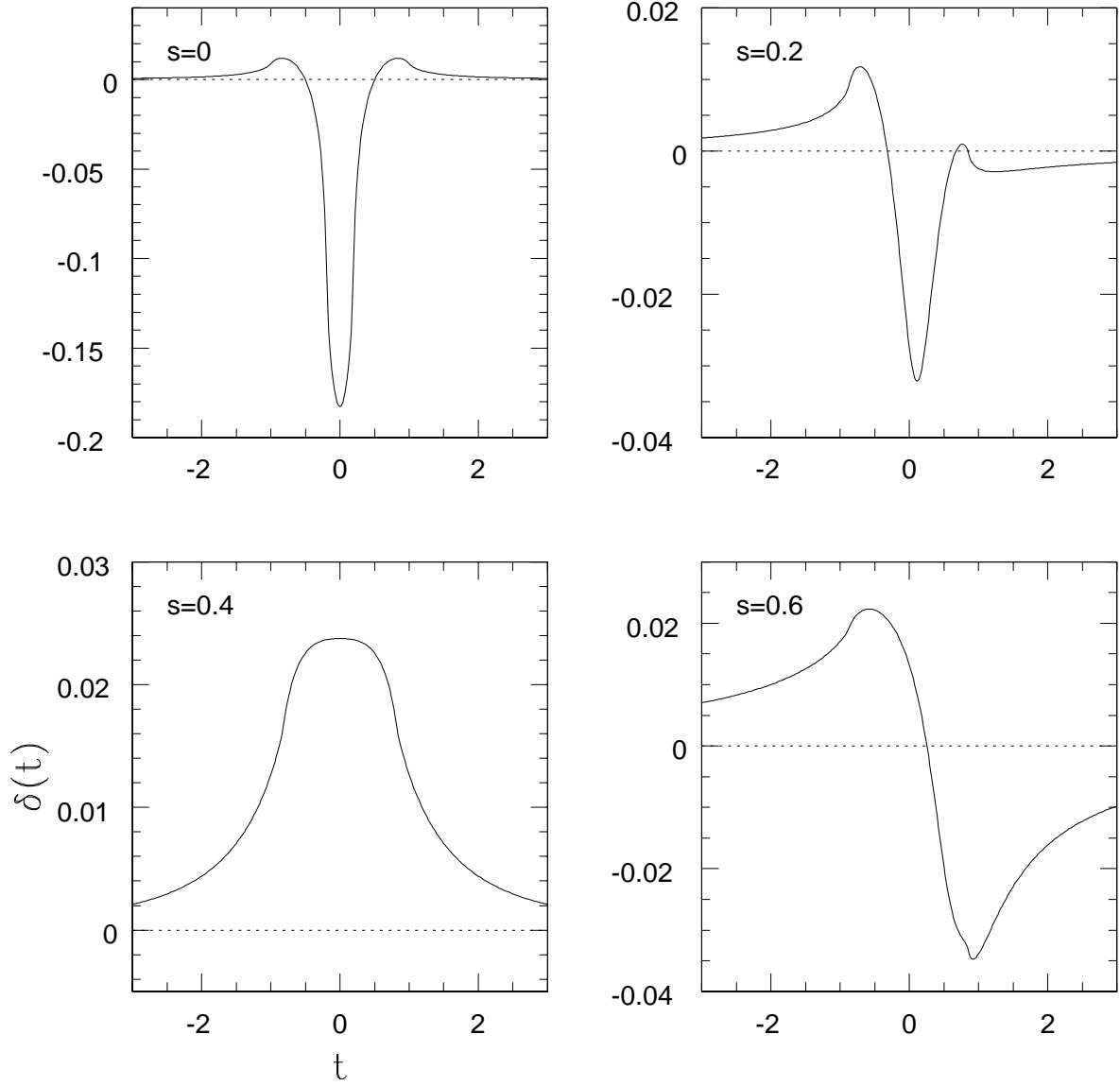}
\caption{Relative light curve deviation $\delta$ as a function of time $t$
(in units of source-radius crossing time) for four lens paths marked in
the corresponding panels of Fig.~\ref{fig:devcontour}. As in the previous
figure, Einstein radius $\epsilon=13.23$, spot radius $r_s=0.2$ and spot
contrast $\mu=10$. The four spot position values $s$ label the upper
left-hand corner of the panels. Lens trajectory parameters for the upper
left-hand panel are $(p,\beta)=(0,0^\circ)$, upper right-hand 
$(0.5,60^\circ)$, lower left-hand $(-0.6,90^\circ)$ and lower right-hand
$(-0.5,-10^\circ)$. Vertical inversions $\delta\rightarrow -\delta$
correspond to deviation curves for a bright spot with $\mu\doteq 0.5$ in
the same configurations.}
\label{fig:devcurve}
\end{figure}

Deviation curves for any particular lensing event with the given spot
geometry can be read off directly from the plots in
Figure~\ref{fig:devcontour}. Examples corresponding to the four lens
paths marked in Figure~\ref{fig:devcontour} are shown in
Figure~\ref{fig:devcurve}. Orienting our coordinate system as in
Figure~\ref{fig:devcontour} with the spot along the positive $x$-axis, we
parametrize the lens trajectory $\vec{r}_L = (x_L,y_L) =
(p\sin\beta+t\cos\beta,-p\cos\beta+t\sin\beta)$, where $t$ is the time in
units of source-radius crossing time measured from closest approach. The
parameter $\beta$ is the angle between the spot position vector $\vec{s}$
and the lens velocity $\dot{\vec{r}}_L$. In this notation the impact
parameter $p$ is given a sign depending on the lens motion - positive if
$\vec{r}_L$ turns counterclockwise, negative if clockwise. The upper
left-hand panel in Figure~\ref{fig:devcurve} corresponds to the maximum
spot-transit effect. The three other panels demonstrate several other
possible light curve deviations.

\begin{figure}[tb]
\plotone{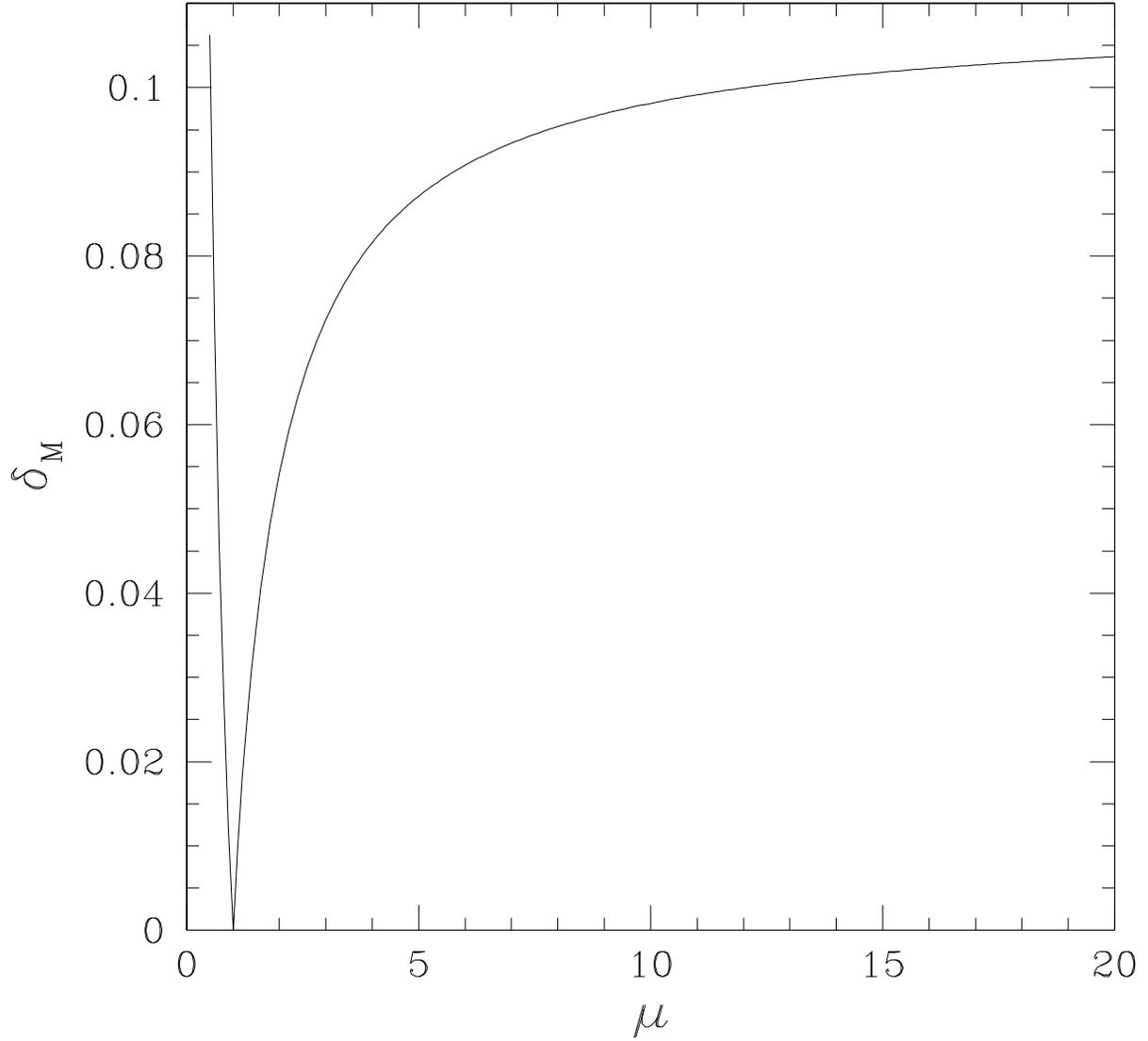}
\caption{Dependence of the maximum deviation $\delta_M$ on spot contrast
$\mu$. This example corresponds to a zero impact parameter transit of a
source with an $r_s=0.1, s=0$ spot by an $\epsilon=13.23$ lens.}
\label{fig:contrast}
\end{figure}

According to equation~(\ref{eq:dev}), the time dependence of the
deviation (i.e., $\vec{r}_L$-dependence) can be separated from the
dependence on the spot contrast ($\mu$ through $B_D$). It follows that
changing the contrast affects only the amplitude of the deviation during
a microlensing event, not its time dependence. To see the change in
amplitude as a function of $\mu$, it is sufficient to look at the change
in the maximum deviation during an event\,\footnote{\,The maximum
deviation is therefore parametrized by $p$ and $\beta$ instead of
$\vec{r}_L$.}, $\delta_M=\max_t|\,\delta [\vec{r}_L(t)]\,|$. The
dependence of $\delta_M$ on spot contrast is shown in
Figure~\ref{fig:contrast}. In this generic example an $\epsilon=13.23$
lens has a zero impact parameter and a $r_s=0.1$ spot is centered on the
source ($s=0$). The dependence is steep for $\mu<5$, but changes only
slowly for $\mu>10$. Values of $\mu$ between 2 and 10 are thought to be
typical of stellar dark spots, roughly corresponding to $\Delta$T values
of 250 to 1000 K, which is at the high end of spots observed in active
stars. In most of the calculations presented here we use $\mu=10$ to
study the maximum spot effect.

In a similar way we can study the dependence on the Einstein radius. We
find that $\delta_M$ grows while $\epsilon<1$, but remains practically
constant for $\epsilon>2$. This saturation is due to the linear
dependence of amplification on $\epsilon$ close to the light curve peak
($|\vec{r}_L|\ll \epsilon$) for sufficiently large $\epsilon$. The ratio
of amplifications in equation~(\ref{eq:dev}) then cancels out the
$\epsilon$-dependence.

\begin{figure}[tb]
\plotone{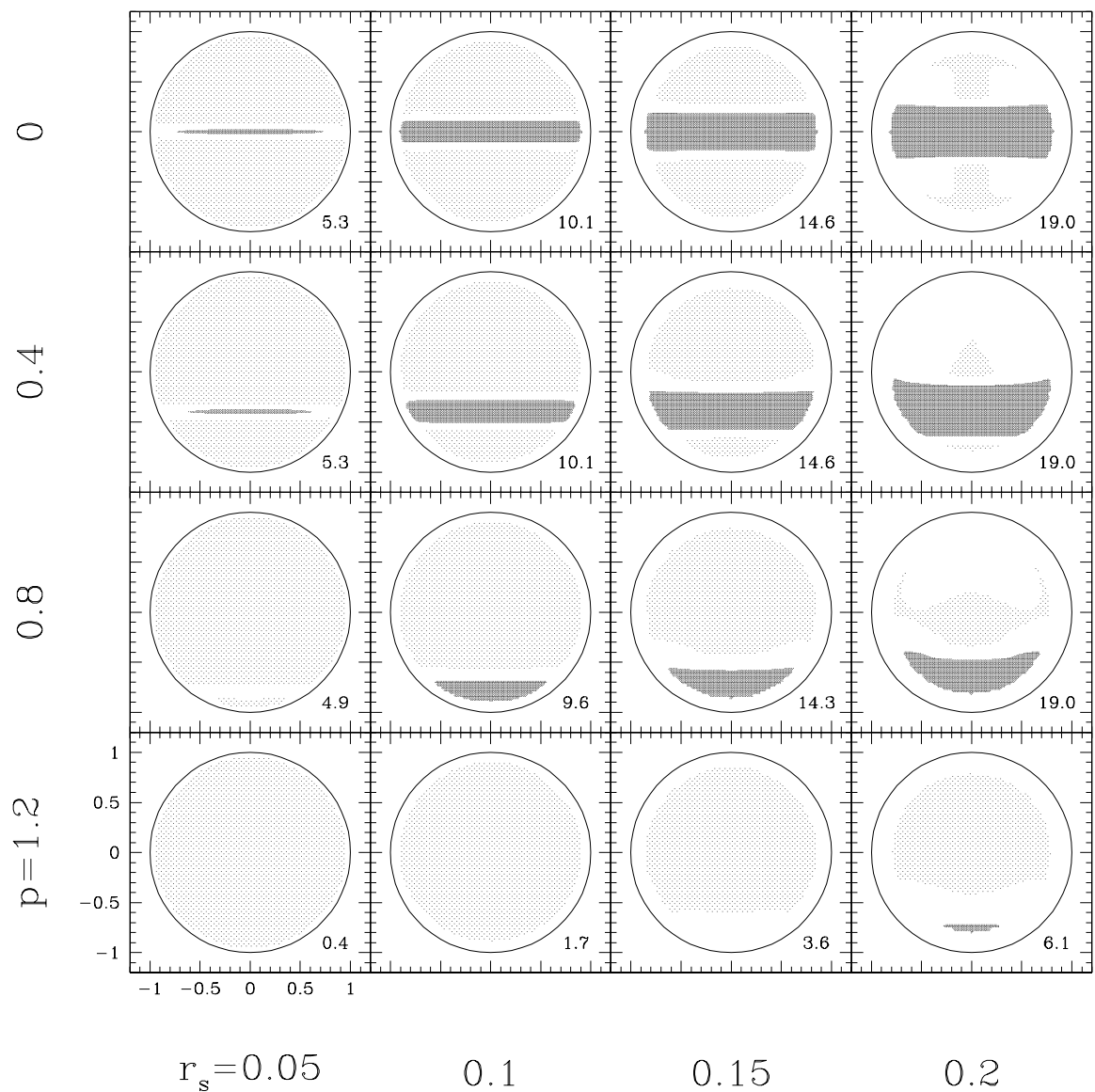}
\caption{Maximum spot effect $\delta_M$ color-coded as a function of spot
position on the projected stellar disk for different impact parameters
$p$ and spot sizes $r_s$. In all cases the $\epsilon=13.23$ lens passes
horizontally in the lower half of the disk. Spot contrast is kept constant
$\mu=10$. Spots centered in the black regions cause $\delta_M>5\%$; those
centered in the gray regions $\delta_M<2\%$. The maximum effect
$\delta_M$ in percent is noted in each of the panels.} 
\label{fig:poseffect}
\end{figure}

The effect of spot size on $\delta_M$ is illustrated by the following two
figures. Figure~\ref{fig:poseffect} demonstrates the detectability of
spots on the projected stellar surface for various combinations of spot
radius and impact parameter. Spots centered in the black regions will
produce a maximum effect higher than 5\%. Those centered in the gray
areas have $\delta_M<2\%$, and thus would be difficult to detect. We can
draw several conclusions for dark spots with sufficient contrast (here
$\mu=10$). As a rule of thumb, small spots with radii $r_s\leq0.15$ could
be detected ($\delta_M>2\%$) if the lens passes within $\sim 1.5\,r_s$ of
the spot center. Larger spots with $r_s\geq0.2$ can be detected over a
large area of the source during source-transit events, and possibly even
marginally during near-transit events (e.g. $p\sim1.2$). As a further
interesting result, the maximum effect a spot of radius $r_s$ (within the
studied range) can have during a transit event is numerically roughly
$\delta_M\sim r_s$, irrespective of the actual impact parameter value.  
For example, a spot with radius 0.05 can have a maximum effect of 5\%,
and an $r_s=0.2$ spot can cause a 19\% deviation.

\begin{figure}[tb]
\plotctyri{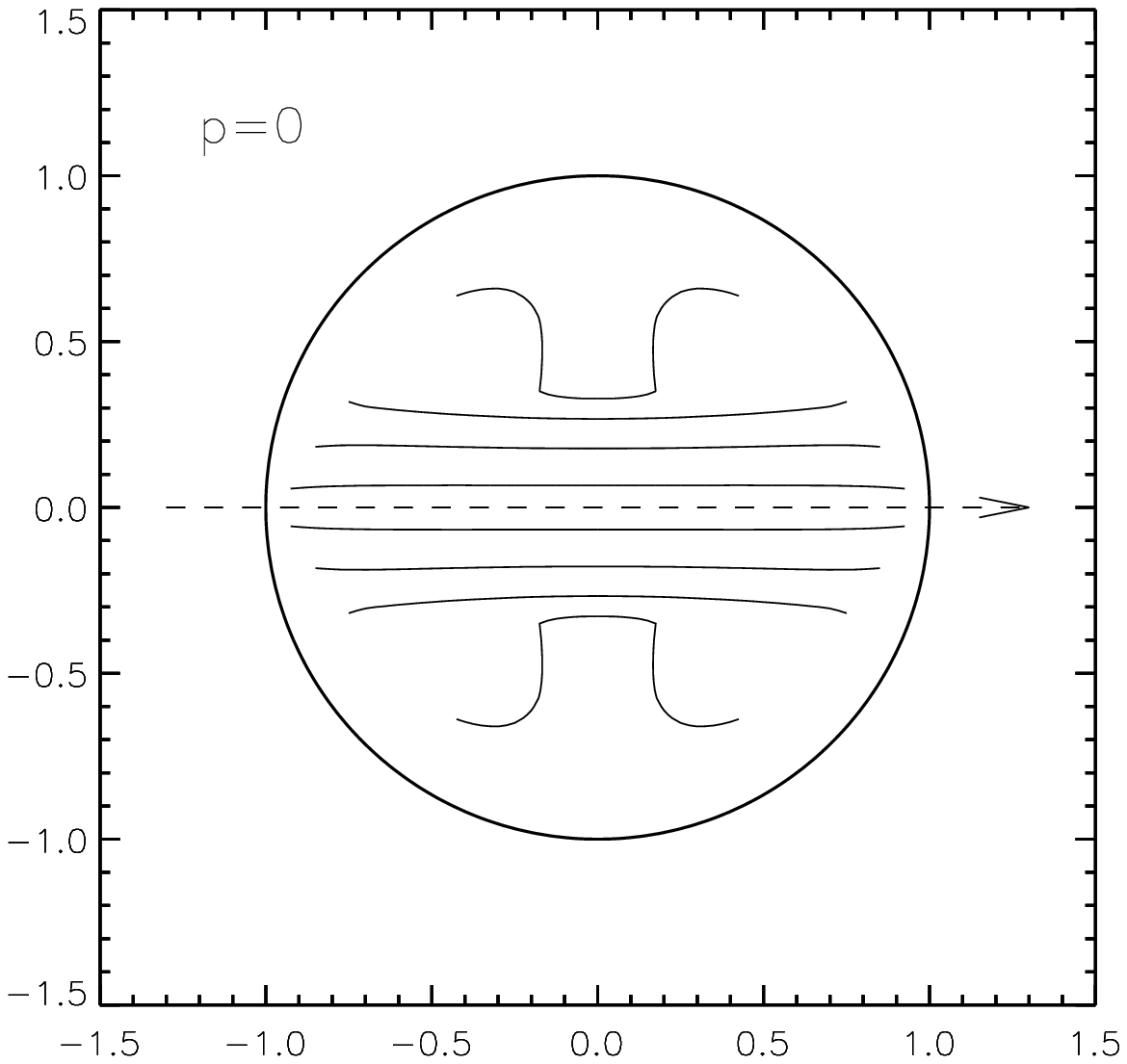}{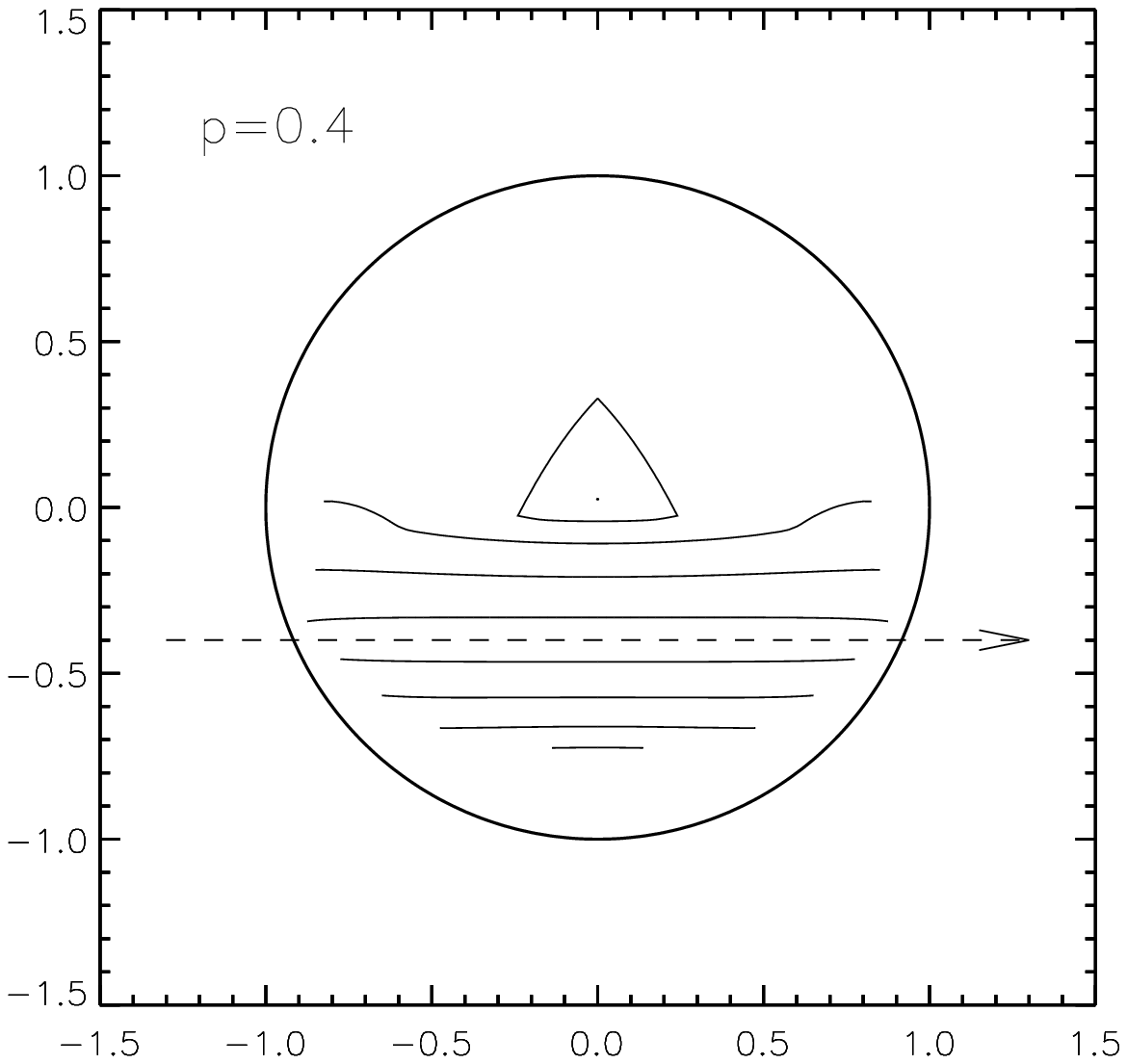}{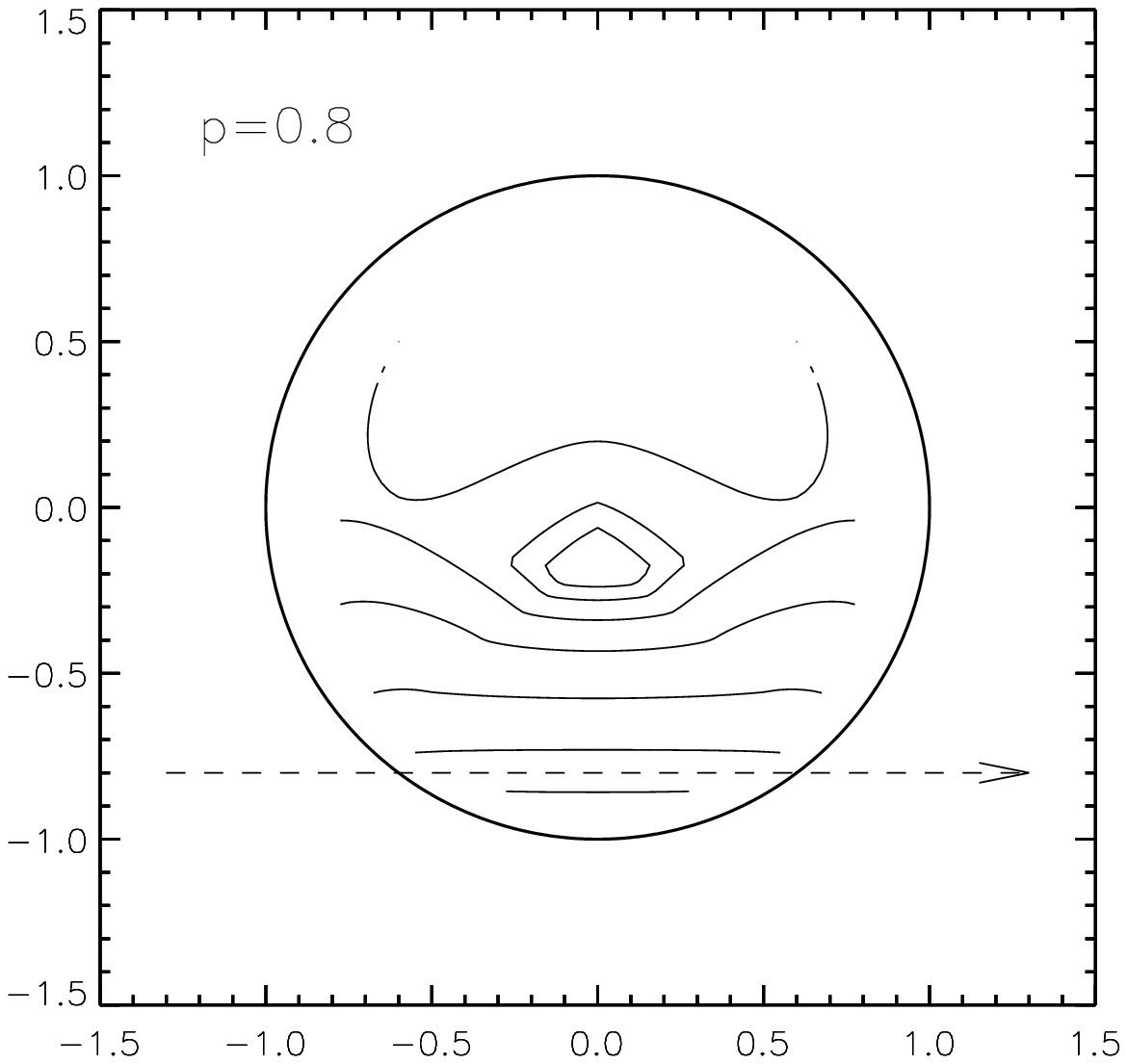}{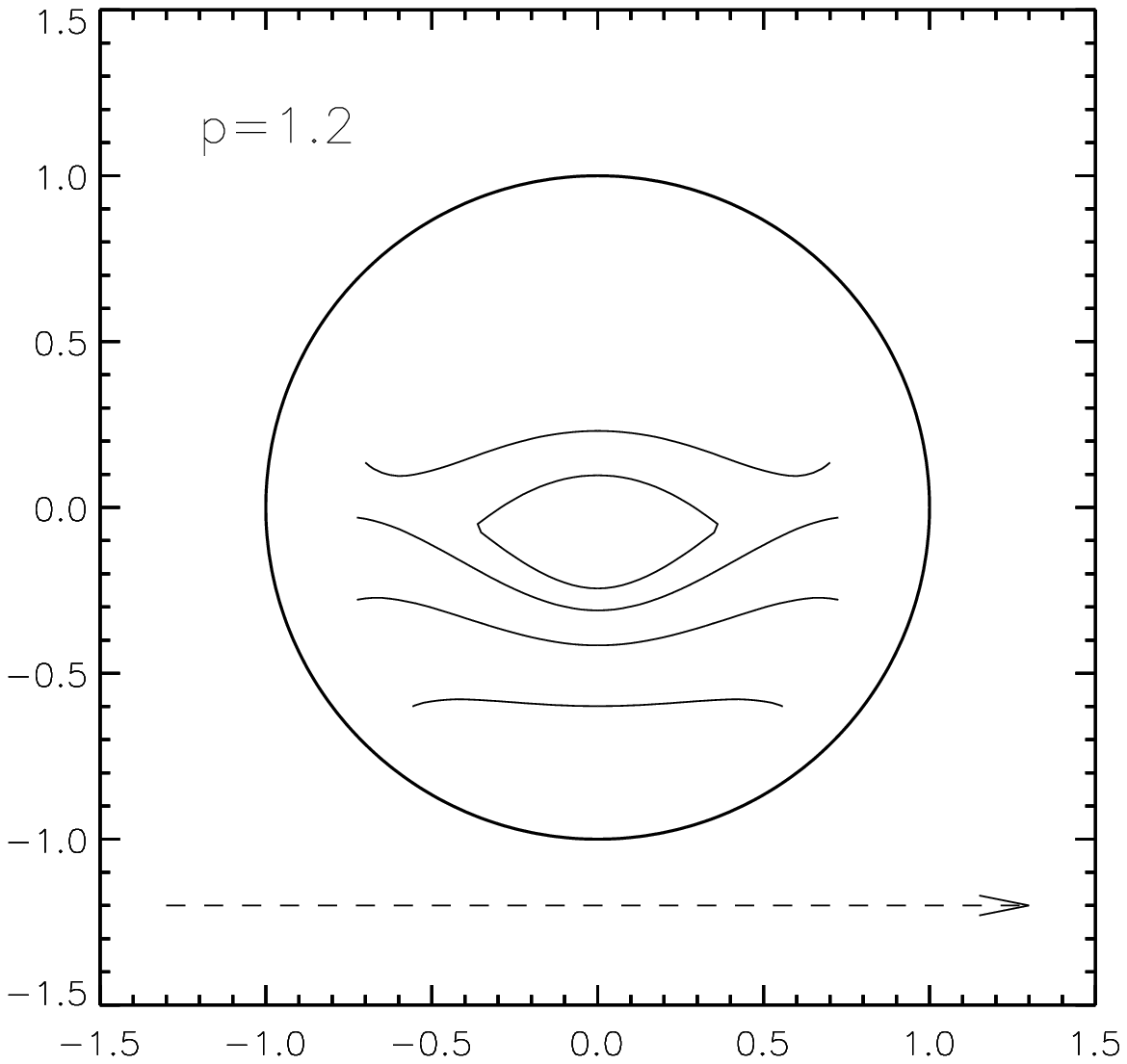}
\caption{Contours of minimum detectable spot size on the projected stellar
disk as a function of spot position in microlensing events with impact
parameters $p=0, 0.4, 0.8, 1.2$ as marked in the panels. Detectability
condition used here is $\delta_M>2\%$. Dashed arrows mark the lens
trajectories. Contour values increase away from the lens path on both 
sides with $0.05$ spacing - for $p=0, 0.4$ values range from $r_s=0.05$ to
$r_s=0.2$; for $p=0.8$ from $r_s=0.05$ to $r_s=0.3$ ({\em inner closed
contour}); for $p=1.2$ from $r_s=0.15$ to $r_s=0.3$ ({\em closed
contour}). As in Fig.~\ref{fig:poseffect}, $\epsilon=13.23$ and
$\mu=10$.}  
\label{fig:minsize}
\end{figure}

Figure~\ref{fig:minsize} is closely related to Figure~\ref{fig:poseffect}.
For a fixed impact parameter we plotted contours of the minimum radius of
a spot (centered at the particular position) necessary to be detectable
($\delta_M>2\%$). As hinted above, during any transit event a spot with
$r_s\simeq0.3$ located practically anywhere on the projected surface of
the source will produce a detectable signature on the light curve.

Turning to the case of a bright spot, we can use the same approach as
above with a negative decrement $B_D$ in formula~(\ref{eq:dev}),
corresponding to contrast parameter $\mu<1$. As noted earlier, a change
in the contrast affects only the amplitude but not the time dependence of
the deviation. The only difference for a bright spot is a change in sign
of the deviation due to negative $B_D$. Therefore the geometry of the
contour plots in Figure~\ref{fig:devcontour} remains the same, only the
contour values and poles are changed. The maximum deviation at the
position of the bright spot is now positive, the weaker opposite peak is
negative. Changing the sign of the deviation in Figure~\ref{fig:devcurve}
in fact gives us deviation curves for a bright spot in the same geometry
with $\mu\doteq 0.5$ instead of $\mu=10$. This correspondence can be seen
from Figure~\ref{fig:contrast}, where both these values have a same
maximum effect $\delta_M$. Unlike in the case of a dark spot, there is
mathematically no upper limit on the relative effect of a bright spot. A
very bright spot would achieve high magnification and dominate the light
curve, acting as an individual source with radius $r_s$.

\section{Change in Spectral Line Profiles}

Studying spectral effects requires computing light curves for a large set
of wavelengths simultaneously. Changes in the observed spectrum of a
spotless source star due to microlensing are described in HSL. Most
individual absorption lines respond in a generic way - they appear less
prominent when the lens is crossing the limb of the source and become
more prominent if the lens approaches the center of the source. The
effect can be measured by the corresponding change in the equivalent
width of the line.

The use of sensitive spectral lines can maximize the search for spots and
active regions on the surfaces of microlensed stars (Sasselov~1997).
Similar techniques are widely known and used in the direct study of the
Sun. One example is observing the bandhead of the CH radical at 430.5~nm,
which provides very high contrast to surface structure (Berger
et~al.~1995). This method will require spectroscopy of the microlensing
event, but could be very rewarding.

For red giants such as the M95-30 source, the H$\alpha$ line will be
sensitive to active regions on the surface (which often, but not always,
accompany spots). To demonstrate the effect, we computed the H$\alpha$
profile using a five-level non-LTE solution for the hydrogen atom in a
giant atmosphere, as described in HSL. We use the same source model as in
the previous calculations (T=3750 K, $\log g$=0.5), for the active region
($\Delta$T$\sim$800 K) we use the line profile of a $\log g$=2 giant with
a chromosphere similar to that of $\beta$ Gem.

\begin{figure}[tb]
\plotone{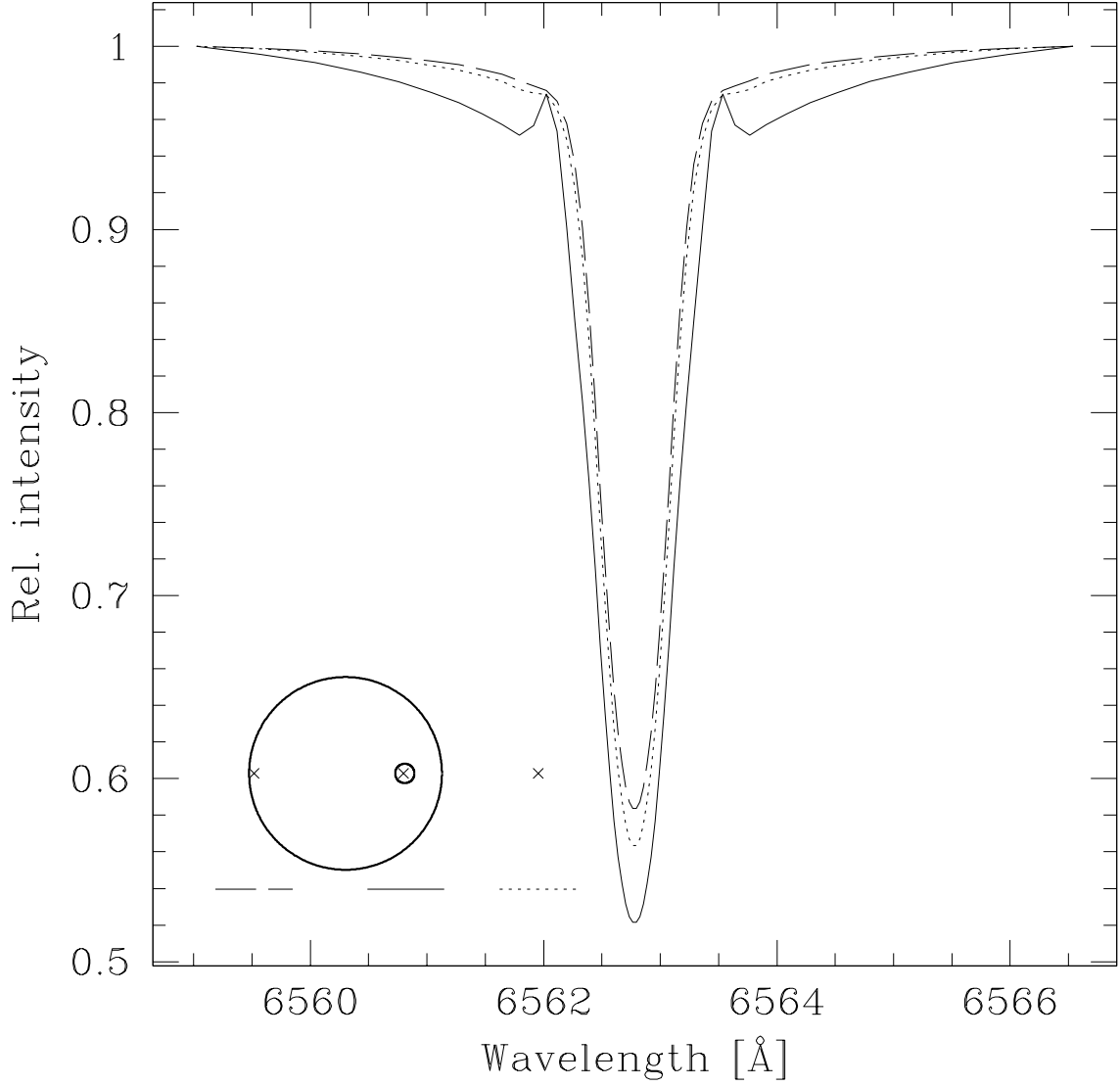}
\caption{H$\alpha$ line profiles of a microlensed star with an active
region for different lens positions. The bold sketch in the lower left
illustrates the stellar disk with an active region of radius $r_s=0.1$.
The three lens positions are marked by crosses. Einstein radius 
$\epsilon=13.23$.}
\label{fig:haspot}
\end{figure}

In Figure~\ref{fig:haspot} we show a time sequence of the changing
profile of H$\alpha$ distorted by an $r_s=0.1$ active region on the
surface of the star. In the calculation we used the M95-30 Einstein
radius and a zero impact parameter. The presence of the H$\alpha$-bright
region leads to a noticeable change in the line profile; in its absence
the change is considerably weaker. In this particular case, more
pronounced wings as well as wing emission can be seen when the lens
passes near the active region.

\section{Discussion}

The small spot model used in this work has obvious limitations. For
example, the constant brightness decrement assumption is not adequate
close to the limb, and spots can have various shapes and brightness
structure (umbra, penumbra). However, most of these problems are not
significant for sufficiently small spots and will not change the general
character of the obtained results. The dependence of the deviation on the
spotless brightness profile $B(r)$ was also neglected in the study
(except for its effect on the contrast $\mu$). According to
equation~(\ref{eq:dev}), it can be expected to have a weak effect on the
amplitude, and due to the spotless amplification in
equation~(\ref{eq:starampl}) an even weaker effect on the time dependence
of the amplification deviation in a microlensing event.

More importantly, it should be noted that the deviations computed in this
paper are deviations from the light curve of the underlying spotless
source in the same lensing geometry. This is not necessarily the best-fit
spotless light curve for the given event. In practice, this will limit
the range of the marginally detectable events with $\delta_M\sim2\%$.
Good photometry and spectroscopy combined with an adequate model
atmosphere for the source star can reduce this problem.

The range of detectability will also be reduced if we consider the
duration of the observable effect. If this effect occurs over too short a
period, it could easily pass undetected. Source-crossing times in
microlensing transit events can reach several days ($\sim3.5$ days in
M95-30 with $p\sim0.7$; corresponding source-radius crossing time
$\sim2.5$ days). As seen from Figures~\ref{fig:devcontour} and
\ref{fig:devcurve}, an effect $\delta>2\%$ can then last hours to days.
Dense light curve sampling during any transit event can therefore lead to
detections or at least provide good constraints on the presence of spots
on the source star. Note that these timescales are too short to expect
effects due to intrinsic changes in the spots or their significant motion
in the case of red giant sources, which have typically slow rotation
speeds. These effects should be considered only in long timescale events
with smaller sources - events with an inherently low probability. In such
events one has to additionally consider the foreshortening of a spot as
it moves towards the limb of the source.

The source star can be expected to have more than just a single spot. The
lensed flux is linear in $B(\vec{r}\,)$; therefore, it can be split again
into terms corresponding to individual spots and the underlying spotless
source. An analysis similar to the one in this paper can then be
performed. The single spot case provides helpful insight into the general
case, even though the relative amplification deviation is not a linear
combination of individual spot terms.

The presence of spots on microlensed stars (e.g., red giants) could
complicate the interpretation of light curves that may be distorted due
to a planetary companion of the lens (Gaudi \& Gould~1997, Gaudi \&
Sackett~1999). A dark spot could be confused with the effect of a planet
perturbing the minor image of the source, while a bright spot ($\mu<1$)
can resemble a major image perturbation (see Figure~\ref{fig:devcurve}).
However, the spot effect is always localized near the peak region of the
light curve, which is itself affected by the finite source size.  
Deviations due to planetary microlensing are usually expected as
perturbations offset from the peak of a simple point-source light curve.
In general it would be sufficient to look for signatures of limb-crossing
during the event, by photometry in two or more spectral bands or by
spectroscopy. Any color effect observable in a planetary microlensing
event will occur within the period of the amplification deviation effect
(Gaudi \& Gould~1997). In the case of a spotted source, the color effect
due to the spot will be preceded and followed by color effects due to
limb crossing (see HSL).

High-magnification planetary microlensing events, in which the source
crosses the perturbed caustic near the primary lens (Griest \&
Safizadeh~1998), can prove to be more difficult to distinguish, as they
can also have a similar limb-crossing signature for a sufficiently large
source. However, in such cases there will be no prominent perturbations
between the two limb crossings, which can at least eliminate confusion
with direct spot transits.

\section{Summary}

Stellar spots can be detected by observations of source-transit
microlensing events. The amplification deviation due to the spot can be
positive as well negative, depending on the relative configuration of the
lens, source, and spot. In the small spot case ($r_s\lesssim0.2$) studied
here, we find that dark spots with radii $r_s\lesssim0.15$ on the
projected stellar disk can cause deviations $\delta_M>2\%$ if the lens
passes within $1.5\,r_s$ of the spot center. Larger spots with
$r_s\sim0.2$ can be detected over a large area of the surface of the
source during any transit event, in some cases even in near-transit
events. Numerically we find that the maximum effect of a dark spot with
sufficient contrast is roughly equal to the fractional spot radius $r_s$,
when the spot is directly crossed. On the other hand a very bright spot
can dominate the shape of the light curve. The obtained results on the
relative amplification deviation are largely independent of the Einstein
radius of the lens in the range $\epsilon>2$; most microlensing events
toward the Galactic bulge fall well within this range. The presence of
spots and especially active regions can also be detected efficiently by
observing the changing profiles of sensitive spectral lines during
microlensing events with a small impact parameter.

Light curves due to sources with spots can resemble in some cases the
effect of a low-mass companion of the lens. Good photometry and
spectroscopy will suffice to distinguish the two in most cases by
detecting additional limb-crossing effects in the case of a spotted star.
Currently operating microlensing follow-up projects such as the Probing
Lensing Anomalies NETwork (PLANET; Albrow et~al.~1998) and Global
Microlensing Alert Network (GMAN; Becker et~al.~1997) can perform
high-precision photometry with high sampling rates; both are sensitive
enough to put constraints on the presence of spots in future
source-transit events. As a result, over a few observing seasons
statistical evidence for spots on red giants could be obtained, making an
important contribution to our theoretical understanding of red giant
atmospheres.

\acknowledgements
We would like to thank Avi Loeb for stimulating discussions and the 
anonymous referee for helpful suggestions on the manuscript.

\clearpage

\end{document}